\documentstyle[prb,aps,epsf,twocolumn]{revtex}

\begin{document}
\draft

\twocolumn[\hsize\textwidth\columnwidth\hsize\csname @twocolumnfalse\endcsname

\title{\bf Diffusion of gold nanoclusters on graphite}

\author{Laurent J.\ Lewis,\cite{byline1}}
\address{D{\'e}partement de Physique et Groupe de Recherche en Physique et
Technologie des Couches Minces (GCM), Universit{\'e} de Montr{\'e}al, Case
Postale 6128, Succursale Centre-Ville, Montr{\'e}al, Qu{\'e}bec, Canada H3C 3J7}

\author{Pablo Jensen,\cite{byline2} Nicolas Combe,\cite{byline3}
and Jean-Louis Barrat\cite{byline4}}
\address{D\'epartement de Physique des Mat\'eriaux, Universit\'e Claude-Bernard
Lyon-I, CNRS UMR 5586, 69622 Villeurbanne C\'edex, France}

\date{\today}

\maketitle

\begin{center}
Submitted to {\em Physical Review B} \\
\end{center}

\begin{abstract}

We present a detailed molecular-dynamics study of the diffusion and
coalescence of large (249-atom) gold clusters on graphite surfaces. The
diffusivity of monoclusters is found to be comparable to that for single
adatoms. Likewise, and even more important, cluster dimers are also found to
diffuse at a rate which is comparable to that for adatoms and monoclusters.
As a consequence, large islands formed by cluster aggregation are also
expected to be mobile. Using kinetic Monte Carlo simulations, and assuming a
proper scaling law for the dependence on size of the diffusivity of large
clusters, we find that islands consisting of as many as 100 monoclusters
should exhibit significant mobility. This result has profound implications
for the morphology of cluster-assembled materials.

\end{abstract}
%    36.40.Sx Diffusion and dynamics of clusters
%    61.46.+w Clusters, nanoparticles, and nanocrystalline materials
%    68.35.Fx Diffusion; interface formation
%    07.05.Tp Computer modeling and simulation
%    81.05.Ys Nanophase materials

\vskip2pc]

\narrowtext

\section{Introduction}
\label{intro}

Nanometer-size clusters --- or simply nanoclusters --- are intrinsically
different from bulk materials.\cite{clufree,science} Yet, understanding of
several of their most fundamental physical properties is just beginning to
emerge (see for instance Refs.\
\onlinecite{nie94,rot94,uzi96,lew97,del97,cha97,uzi99}), thanks largely to
rapid progresses in the technology of fabrication and analysis, but also
considerable advances in computational tools and methodology.

It has recently been demonstrated\cite{paillard,perez,pablo} that depositing
clusters (rather than single atoms) on surfaces allow the fabrication of
interesting nanostructured materials whose properties can be tailored to
specific technological applications, e.g., micro-electronic, optoelectronic,
and magnetic devices.\cite{nanos} If single-atom deposition is used, the
nanostructures have to be grown directly on the substrate through diffusion
and agregation, which depends in a detailed (and in general very complicated)
manner on the interactions between surface atoms and adatoms. By contrast,
for cluster deposition, the clusters are prepared before they hit the
surface, giving considerably more flexibility\cite{size} in assembling or
organizing clusters for particular applications. It has been shown, for
instance, that by changing the mean size of the incident carbon clusters, it
was possible to modify the structure of the resulting carbon film from
graphitic to diamond-like.\cite{paillard} This however requires that
sufficient control over the cluster deposition and subsequent growth process
be achieved.\cite{perez,pablo}

Diffusion evidently plays a central role in the fabrication of thin films and
self-organized structures by cluster deposition. It has been demonstrated
experimentally that gold or antimony clusters diffuse on graphite surfaces at
a surprisingly high rate of about $10^{-8}$ cm$^2$/s at room
temperature,\cite{PJLB} quite comparable to the rates that can be achieved by
single atoms in similar conditions. This was confirmed theoretically by
Deltour {\em et al.} using molecular-dynamics simulations:\cite{del97}
clusters consisting of particles which are incommensurate with the substrate
exhibit very rapid diffusion. The cluster diffuses ``as a whole'', and its
path is akin to a Brownian motion induced by the internal vibrations of the
clusters and/or the vibrations of the substrate. This is in striking contrast
with other cluster diffusion mechanisms, whereby the motion results from a
combination of single-atom processes, such as evaporation-condensation, edge
diffusion, etc. The latter mechanisms are more appropriate to clusters which
are in epitaxy with the surface, and are likely not significant in cases
where the mismatch is large and/or the substrate-cluster interactions are
weak, such as in Refs.\ \onlinecite{PJLB} and \onlinecite{kern} (see Ref.\
\onlinecite{pablo} for a review).

In the present paper, we re-examine the problem of cluster diffusion in the
cluster-substrate-mismatched case, now using a much more accurate model:
indeed, in the work of Deltour {\em et al.},\cite{del97} cluster-cluster,
cluster-substrate and substrate-substrate interactions were all assumed to be
of the Lennard-Jones form, which cannot be expected to correctly describe
``real materials''. Here, we consider a simple, but realistic model for the
diffusion of gold clusters on a graphite surface (HOPG). We are concerned
with gold because it has been the object of several experimental
studies,\cite{pablo,fluelli,borel,ss} but also because realistic
semi-empirical, many-body potentials are available for this material. The
energetics of gold atoms is described in terms of the embedded-atom-method
(EAM),\cite{fbd} while carbon atoms are assumed to interact via Tersoff
potentials;\cite{tersoff} the (weak) interactions between gold and carbon
atoms are modeled with a simple Lennard-Jones potential. A comparable model
was used recently by Luedtke and Landman to study the anomalous diffusion of
a gold nanocluster on graphite;\cite{uzi99} diffusion was found to proceed
via a stick-slip mechanism, resulting in an apparent L\'evy-flight type of
motion. In the present work, we examine closely the variations with
temperature of the rate of diffusion, as well as the microscopics of cluster
dimers (diclusters).

We find the diffusivity of monoclusters to be entirely comparable to that for
single adatoms. Likewise, and most important, diclusters are also found to
diffuse at a rate which is comparable to that for adatoms and monoclusters.
It is therefore expected that large islands, formed by the aggregation of
many clusters, should also be mobile. Based on this observation, we carried
out kinetic Monte Carlo simulations of island diffusion and coalescence
assuming a proper scaling law for the dependence on size of the diffusivity
of large clusters. We find that islands consisting of as many as 100
monoclusters exhibit significant mobility; this is consistent with the
observation on graphite of large (200 monoclusters) gold islands. The
morphology of cluster-assembled materials is profoundly affected by the
mobility of multi-cluster islands.

\section{Computational details}
\label{comp_det}

Diffusion coefficients for clusters can only be obtained at the expense of
very long MD runs: there exists numerous possible diffusion paths, and there
is therefore not a single energy barrier (and prefactor) characterizing the
dynamics. These systems, further, do not lend themselves readily to
accelerated MD algorithms.\cite{voter,fichthorn} Brute-force simulations --
long enough for statistically-significant data to be collected -- therefore
appear to be the only avenue. This rules out {\em ab initio} methods, which
can only deal with very small systems (a few tens of atoms) over limited
timescales (tens of picoseconds at best): empirical or semi-empirical
potentials {\em must} be employed.

As mentioned above, we describe here the interactions between Au particles
using the embedded-atom method (EAM),\cite{fbd} an $n$-body potential with
proven ability to describe reliably various static and dynamic properties of
transition and noble metals in either bulk or surface
configurations.\cite{eam-review} The model is ``semi-empirical'' in the sense
that it approaches the total-energy problem from a local electron-density
viewpoint, but using a functional form with parameters fitted to experiment
(equilibrium lattice constant, sublimation energy, bulk modulus, elastic
constants, etc.).

The interactions between C atoms are modelled using the Tersoff
potential,\cite{tersoff} an empirical $n$-body potential which accounts well
for various conformations of carbon. The Tersoff potential for carbon is
truncated at 2.10 \AA, which turns out to be smaller than the inter-plane
distance in graphite, 3.35 \AA. Thus, within this model, there are no
interactions between neighbouring graphite planes. This is of course an
approximation, but not a bad one since basal planes in graphite are known to
interact weekly. (This is why it is a good lubricant!). A pleasant
consequence of this is that the substrate can be assumed to consist of a
single and only layer, thus reducing formidably the (nevertheless very heavy)
computational load of the calculations.

Last --- and most problematic --- is the Au-C interaction, for which no
simple (empirical or semi-empirical) model is to our knowledge available. One
way of determining this would be to fit an {\em ab initio} database to a
proper, manageable functional potential. However, since Au-C pairs conform in
so many different ways in the present problem, this appears to be a hopeless
task, not worth the effort in view of the other approximations we have to
live with. We therefore improvised this interaction a little bit and took it
to be of the Lennard-Jones form, with $\sigma = 2.74$ \AA\ and $\epsilon = $
0.022 eV, truncated at 4.50 \AA. The parameters were determined rather
loosely from various two-body models for Ag-C and Pt-C
interactions.\cite{loose} Overall, we expect our model to provide a {\em
qualitatively correct} description of the system, {\em realistic} in that the
most important physical characteristics are well taken into account. It is
however not expected to provide a quantitatively {\em precise} account of the
particular system under consideration, but should be relevant to several
types of metallic clusters which bind weekly to graphite.

We consider here gold nanoclusters comprising 249 atoms, a size which is
close to that of clusters deposited in the experiments.\cite{pablo,ss} The
graphite layer has dimensions 66.15$\times$63.65 \AA$^2$ and contains 1500
atoms. Calculations were carried out for several temperatures in the range
400--900 K. It should be noted that a free-standing 249-atom Au cluster melts
at about 650 K in this model.\cite{lew97} This temperature is not affected in
a significant manner by the graphite substrate as the interaction between Au
and graphite-C atoms are weak. However, the dynamics of the cluster is
expected to be different in the high-temperature molten state and the
low-temperature solid state. All simulations were microcanonical, except for
the initial thermalisation period at each temperature; no drift in the
temperature was observed.

Simulations were carried out in most cases using a fully dynamical substrate.
In two cases, one for the single cluster and the other for the dicluster,
extremely long runs using a static (frozen) substrate were performed: it has
been found by Deltour {\em et al.}\cite{del97} that diffusion is
quantitatively similar on both substrates (however, see Section \ref{static}
below). The equations of motion were integrated using the velocity form of
the Verlet algorithm with a timestep of 1.0 and 2.5 fs for dynamic and static
substrates, respectively.\cite{MD} (Carbon being a light atom, a smaller
timestep is needed in order to properly describe the motion). The
dynamic-substrate simulations ran between 10 and 14 million timesteps
(depending on temperature), i.e., 10--14 ns. The static-substrate simulation
for the monocluster, in comparison, ran for a total of 50 million timesteps,
i.e., a very respectable 125 ns = 0.125 $\mu$s; the corresponding dicluster
simulation ran for 75 ns. All calculations were performed using the program
{\tt groF}, a general-purpose MD code for bulk and surfaces developed by one
of the authors (LJL).

\section{Results}
\label{results}

\subsection{Dynamic-substrate simulations}
\label{dynamic}

\begin{figure}
\epsfxsize=8cm
\epsfbox{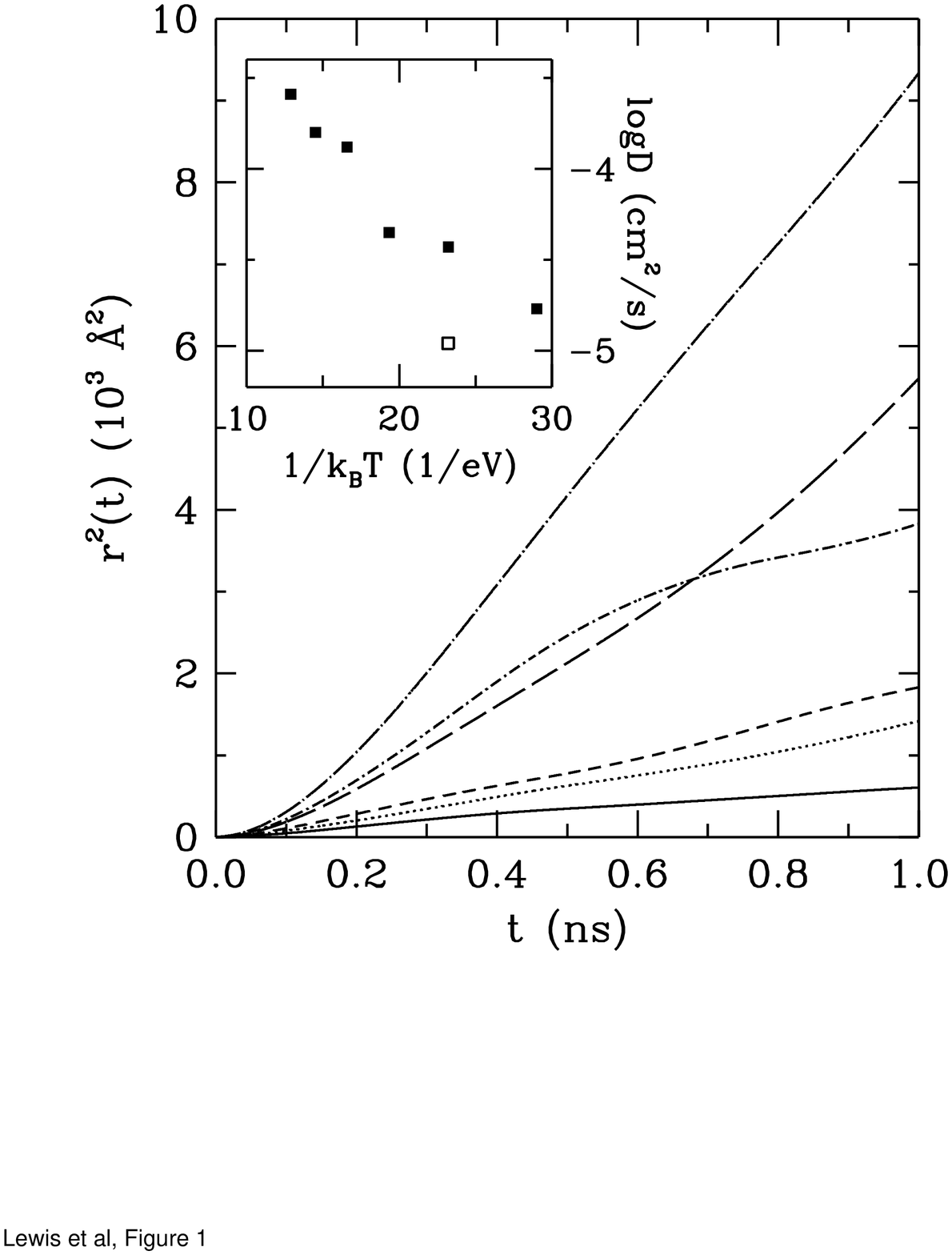}
%\vspace*{-5cm}
\caption{
Main figure: Time-averaged mean-square displacements for the cluster's
center-of-mass at the various temperatures investigated, namely, from bottom
to top, 400, 500, 600, 700, 800 and 900 K. (The 700 and 800 K curves are
inverted for $t > 0.7$ ns). Inset: Arrhenius plot of the diffusion
coefficient. The open square at 500 K ($1/k_BT = 23.2$ eV$^{-1}$) is the
result for the frozen substrate.
\label{fig_diff}
}
\end{figure}

We first discuss diffusion on a dynamic substrate, i.e., with {\em all} parts
of the system explicitly dealt with in the MD simulations. Fig.\
\ref{fig_diff} gives the (time-averaged) mean-square displacements (MSD's) of
the cluster's center-of-mass at the various temperatures investigated, which
will be used to calculate the diffusion constant, $D =
\lim_{t\rightarrow\infty} r^2(t)/4t$. As indicated above, the simulations
extend over 10--14 ns, but the MSD's are only shown for a maximum correlation
time of one ns in order to ``ensure'' statistical reliability. It is evident
(e.g., upon comparing the results at 700 and 800 K) that the diffusion
coefficients that can be extracted from these plots will carry a sizeable
error bar. Nevertheless, it is certainly the case that (i) diffusion is {\em
very significant} and (ii) it increases rapidly with temperature.

There is no evidence from these plots that the MSD's obey a non-linear power
law behaviour (i.e., that the cluster undergoes superdiffusion) which could
be associated with ``L\'evy flights'': the statistical accuracy of the data is
simply not sufficient to draw any conclusions. The cluster does however
undergo long jumps during the course of its motion. We will return to this
point below when we discuss diffusion on a frozen substrate.

In lack of a better description of the long-time behaviour of the diffusion
process, we simply assume that $r^2(t) \rightarrow 4Dt$ as $t$ gets large.
The resulting diffusion coefficients are plotted in the manner of Arrhenius,
i.e., $\log D$ {\em vs} $1/k_BT$, in the inset of Fig.\ \ref{fig_diff}. If
the process were truly Arrhenius, all points would fall on a single straight
line. This is evidently not the case here. Though we could probably go ahead
and fit the data to a straight line, attributing the discrepancies to
statistical error, there is probably a natural explanation for the ``break''
that a sharp eye can observe between 600 and 700 K: As noted above, the free
Au$_{\rm 249}$ cluster melts at about 650 K in the EAM model.\cite{lew97} The
presence of the substrate raises the melting point, but very little since the
interactions between the cluster and the graphite surface are small. Thus,
the cluster is solid at the lowest temperatures (400, 500 and 600 K), but
liquid above (700, 800 and 900 K). The statistics are evidently insufficient
to allow firm conclusions to be drawn; there nevertheless appears to be a
discontinuity near the cluster melting point temperature, with activation
energies on either side of about 0.05 eV. We discuss in Section \ref{growth}
the implications of these findings on the kinetics of growth.

\subsection{Static-substrate simulations}
\label{static}

The static-substrate simulations, carried out at a single temperature (for
the cluster), {\em viz.}\ 500 K, serve many purposes: (i) re-assess the
equivalence with dynamic-substrate MD runs reported by Deltour {\em et
al.};\cite{del97} (ii) provide accurate statistics for a proper comparison of
the diffusive behaviour of mono- and diclusters; (iii) examine the possible
superdiffusive character of the trajectories.

We focus, first, on a comparison between static- and dynamic-substrate
simulations. As can be appreciated from the MSD's given in the inset of Fig.\
\ref{fig_diff_stat}, there is a rather substantial difference between the two
calculations: for the dynamic substrate at 500 K, the diffusion constant is
$3.71 \times 10^{-5}$ cm$^2$/s, while for the frozen substrate we have $1.09
\times 10^{-5}$ cm$^2$/s. (This value is actually significantly smaller than
that for the dynamic substrate at 400 K --- 100 K lower temperature --- {\em
viz.}\ $1.70 \times 10^{-5}$ cm$^2$/s). Again, statistical uncertainties
cannot be totally excluded to account for this discrepancy, but it is
difficult to imagine that it could explain all of the observed difference
(cf.\ inset to Fig.\ \ref{fig_diff} for a better appreciation of this
difference). The explanation might however be quite simple.

As noted above, the cluster-substrate interactions are weak, and this likely
plays an important role in determining the characteristics of the motion.
Visual inspection of the $x-y$ paths in the two different situations makes it
apparent that the motion has a much stronger ``stick-and-jump'' character on
the frozen substrate than on the dynamic one. On the frozen substrate, further,
the trajectory is more compact on a given timescale. This can in fact be
characterized in a quantitative manner by considering, following Luedtke and
Landman,\cite{uzi99} the function $P_\tau(d)$, which gives the distribution
of displacements of length $d$ over a timescale of $\tau$. The motion is best
characterized using a value of $\tau$ corresponding to the period of
vibration of the cluster in a sticking mode (see below).

The function $P_\tau(d)$ (normalized to unity) is displayed in Fig.\
\ref{fig_p_of_d} for the dynamic substrate at three different temperatures
(400, 500, and 900 K) and for the static substrate at 500 K. The value of
$\tau$ was determined from the frozen-substrate simulations by simply
counting the number of oscillations over a given period of time; we found
$\tau = 20$ ps to within about 10\%. We note that, for the dynamic substrate,
the period of oscillations at 400 K is about 38 ps, while no oscillations can
be found at 500 and higher temperatures, i.e., the sticking mode is absent
above 500 K or so.

The difference between static and dynamic substrates is striking: On the
dynamic surface, $P_\tau(d)$ is a broad featureless distribution, which gets
broader as temperature increases. The maximum of the distribution at low
temperature lies at about 1.6--1.8 \AA\ --- roughly the distance between
equilibrium sites on the graphite surface --- clearly establishing that the
motion proceeds in an quasi continuous manner via ``sliding hops'' to
nearest-neighbours; the hops get longer as temperature increases. On the
static substrate, in contrast, a ``sticky'' vibrational mode, of amplitude
roughly 0.25 \AA, is clearly visible. This is followed by a broad tail which
corresponds, again, to the sliding jumps that are characteristic of the
motion on the dynamic substrate.

Sticking, therefore, is much more likely to take place on the static than on
the dynamic substrate, thereby contributing to decrease the average distance
traveled by the cluster over a given period of time. This conclusion is
however not general: The system under consideration here is perhaps a bit
peculiar in that the cluster-substrate interactions are especially weak. (In
comparison, Luedtke and Landman's $\epsilon$ for the Au-C interaction is
0.01273 eV, even smaller than our own value.) One may conjecture that the
vibrations of the surface are enough, in such cases, to overcome {\em
completely} the barrier opposing diffusion, which might not be true of
systems where the interactions are stronger (as in the case, e.g., of Deltour
{\em et al.}'s simulations, Ref.\ \onlinecite{del97}).

\begin{figure}
\epsfxsize=8cm
\epsfbox{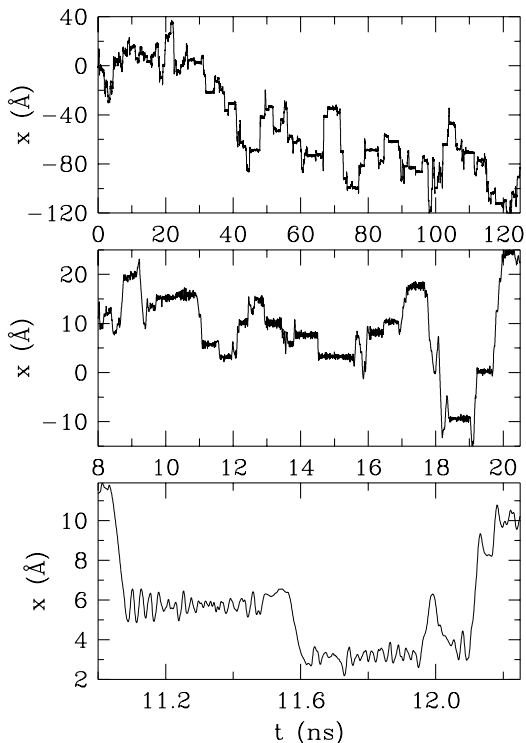}
\vspace*{1.8cm}
\caption{
The $x$ position of the cluster's center of mass at 500 K on a frozen
substrate for three different timescales, showing the apparently self-similar
character of the trajectory.
\label{fig_com}
}
\end{figure}

It also appears that our diffusion data do not cover a timescale long enough
to warrant firm conclusions to be drawn on the possibility that
superdiffusion might be taking place. This is certainly true, as we have seen
above, of the dynamic-substrate simulations, which cover ``only'' 10--14
nanoseconds, but also of the static-substrate simulations (assuming, in view
of the above discussion, that they are relevant to the problem under study),
which extend to 125 ns. Certainly, the position of the cluster's
center-of-mass does exhibit something of a self-similar character, as
reported by Luedtke and Landman\cite{uzi99}, and as can be seen in Fig.\
\ref{fig_com}. Evidently, one cannot trust statistics here over more than a
decade or two in time. One might hope that superdiffusion would be more
apparent in the long-time behaviour of the MSD's. To this effect, we plot in
Fig.\ \ref{fig_diff_stat} $\log r^2(t)$ {\em vs} $\log t$, for a maximum
correlation time of (here) 40 ns;\cite{note} the slope of such a plot is the
diffusivity exponent $\gamma$. The statistical quality of the data decreases
with correlation time, and becomes clearly insufficient over 5 ns or so; the
large dip at about 15 ns can testify. Our best estimate of the slope $\gamma$
at ``large'' (more than $\sim$1 ns) correlation times is anywhere between 0.9
and 1.2, i.e., mild underdiffusion or mild superdiffusion... or no
superdiffusion at all! This is consistent with the value reported by Luedtke
and Landman, who find $\gamma = 1.1$ based on an analysis of sticking and
sliding times. One point worth mentioning is that the
velocity-autocorrelation function for adatom diffusion in the intermediate
and high-friction regimes has been shown to follow a power-law behaviour at
intermediate times; the exponential dependence resumes at very long
times.\cite{tapio}

\subsection{Diclusters}
\label{diclus}

The morphology of films grown by cluster deposition depends critically on the
coefficient of diffusion of monoclusters, as we have just seen, but also,
because clusters aggregate, on the coefficient of diffusion of multiclusters.
From simple geometric arguments, it might be argued that the rate of
diffusion should scale as $N^{-2/3}$, where $N$ is the number of atoms in the
cluster, as was in fact observed by Deltour {\em et al.}\cite{del97} for
Lennard-Jones clusters. However, it can be expected that the morphology of
the films depends, as well, on the {\em shape} of the multi-clusters
following the aggregation of monoclusters, i.e., on the kinetics of
coalescence.

In a previous publication,\cite{lew97} we examined the coalescence of gold
nanoclusters in vacuum and found it to be much slower than predicted by
macroscopic theories. This state of affairs can be attributed to the presence
of facets and edges which constitute barriers to the transport of particles
required for coalescence to take place \cite{epjb}. The ``neck'' between
two particles was however found to form very rapidly. We conjectured that
these conclusions would apply equally well to the particular case of gold
nanoclusters on graphite since the gold-graphite interactions are weak.

\begin{figure}
\epsfxsize=8cm
\epsfbox{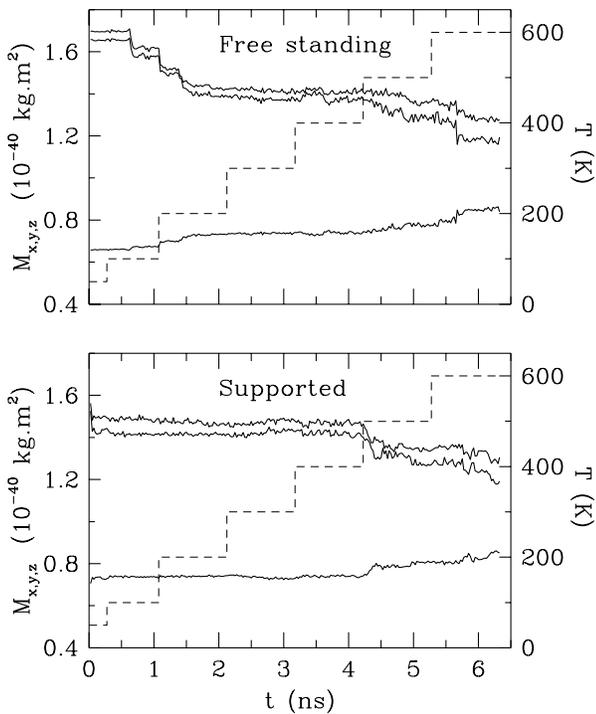}
\vspace*{1.8cm}
\caption{
Evolution with time-temperature of the three moments of inertia (full lines)
of the dicluster, both free-standing and supported on graphite; the
correspondence between temperature and time is indicated by the dashed,
stepwise curve.
\label{fig_coa}
}
\end{figure}

\begin{figure}
\epsfxsize=8cm
\epsfbox{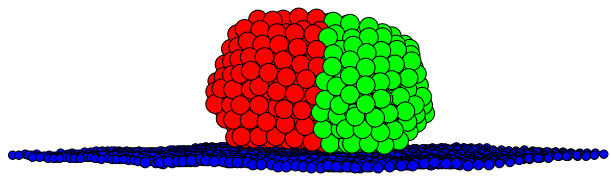}
\vspace*{-3cm}
\caption{
Ball-and stick model of the gold dicluster on the graphite substrate at 200
K, after the neck between the two monoclusters has formed completely. The two
monoclusters are colored differently for ease of visualisation.
\label{fig_dicluster}
}
\end{figure}

We have verified this in the context of the present work: indeed, coalescence
is little affected by the presence of the substrate, as demonstrated in Fig.\
\ref{fig_coa}. We considered both a free-standing and a supported pair of
249-atom gold clusters. Starting at very low temperature (50 K), temperature
was slowly and progressively (stepwise) raised to 600 K. (As noted above, the
249-atom gold cluster melts at about 650 K in this EAM model and we therefore
did not go beyond this point). We plot, in Fig.\ \ref{fig_coa}, the evolution
with time-temperature of the three moments of inertia of the dicluster. Since
the cluster can rotate, the moments of inertia provide a more useful measure
of the shape of the object than, e.g., the radii of gyration.\cite{lew97} A
side view of the dicluster at 200 K, i.e., after the neck between the two
monoclusters has formed completely, is shown in Fig.\ \ref{fig_dicluster}. It
is evident that the dicluster does not wet the surface, and therefore the
substrate plays a relatively minor role in the coalescence process.

As can be seen in Fig.\ \ref{fig_coa}, the behaviour of the free-standing and
supported diclusters are almost identical, except for the initial phase of
coalescence: the supported cluster forms a neck much more rapidly than the
free-standing cluster, presumably because the substrate offers, through some
thermostatic effect, an additional route via which coalescence (by plastic
deformation) can be mediated; it is conceivable also that the substrate
``forces'' the atomic planes from the two clusters to align. We have not
explored these questions further; it remains that the end points of the two
coalescence runs are identical within statistical uncertainty. Thus, again,
coalescence is hampered by the presence of facets and edges; the timescale
for complete coalescence is much longer than predicted by continuous
theories. The shape of islands on the graphite surface will be strongly
affected, and it is also expected that the rate of diffusion will be affected
(since it is determined by the contact area between substrate and cluster).

The MSD of the dicluster (after proper equilibration at 500 K) is displayed
in the inset of Fig.\ \ref{fig_diff_stat}. As mentioned earlier, this was
calculated from a static-substrate run covering 75 ns. The same limitations
as noted above for the monocluster should therefore hold in the present case.
It is a very remarkable (and perhaps even surprising) result that the rate of
diffusion of the dicluster is quite comparable to that of the monocluster,
inasmuch as the frozen-substrate simulations are concerned. (We expect the
diffusion constants on the dynamic substrate to be different --- and larger
--- but in a proportion that would be quite comparable to that found here).
The value of $D = 1.38 \times 10^{-5}$ cm$^2$/s we obtain for the dicluster
is in fact a bit larger than that for the monocluster ($1.09 \times 10^{-5}$
cm$^2$/s). The difference is probably not meaningful; what {\em is}
meaningful, however, is that {the the mono- and the dicluster have comparable
coefficients of diffusion}; this has profound implications on growth, as we
discuss in Section \ref{growth}, below.

\begin{figure}
\epsfxsize=8cm
\epsfbox{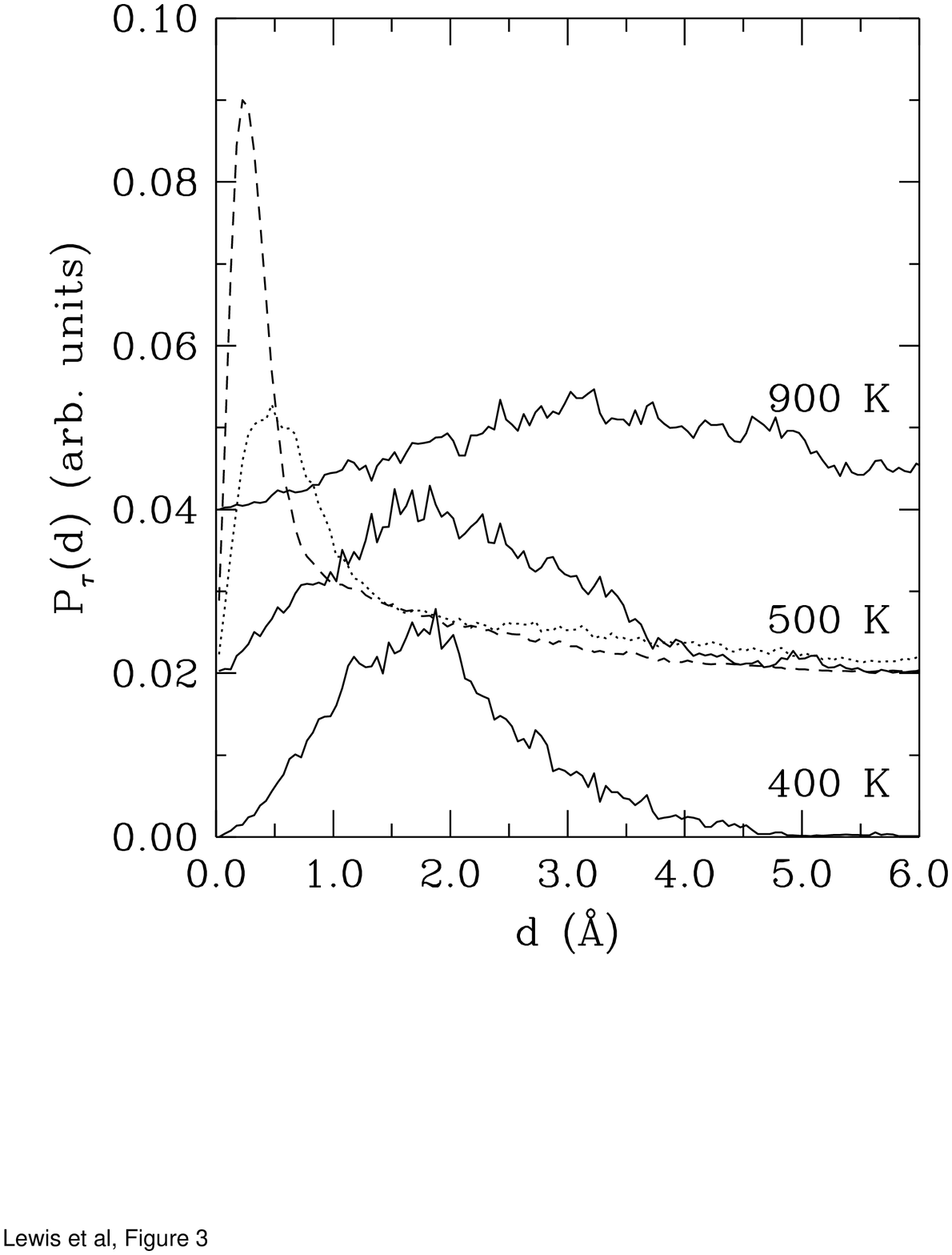}
%\vspace*{-5cm}
\caption{
The function $P_\tau(d)$, which gives the distribution of displacements of
length $d$ over a timescale of $\tau$, at three different temperatures, as
indicated. At 500 K, the three curves correspond to the monocluster on a
dynamic substrate (full line), the monocluster on a static substrate (dashes)
and the dicluster on a static substrate (dots).
\label{fig_p_of_d}
}
\end{figure}

The function $P(d)$ for the dicluster at 500 K is displayed in Fig.\
\ref{fig_p_of_d}; here we estimate that $\tau=40$ ps (vs about 20 ps for the
monocluster). The distribution is quite similar to that found for the single
cluster on the frozen substrate, though broader and shifted to slighlty
larger displacements. This last result is likely due to the fact that, being
larger, the dicluster is not as easily able to accomodate itself with the
substrate as the mo\-no\-clus\-ter; in this sense, it is more loosely bound
to the substrate.

\subsection{Comparison with experimental results}
\label{growth}

Deposition of gold clusters on graphite experiments were carried out in Lyon
recently.\cite{pablo,ss} Several models have been proposed to extract the
microscopic cluster diffusion coefficients from the measured island
densities.\cite{pablo} Of course, in order to provide a meaningful
interpretation of the data, the models must take into account the precise
conditions in which the experiments are performed. In Lyon, for instance, the
flux of clusters is chopped, rather than continuous, and this affects the
kinetics of diffusion and growth considerably.\cite{hache,ncombe} Previous
estimates of the rates of diffusion of Au on graphite, which overlooked this
important detail, are therefore in error. In Ref.\ \onlinecite{ss}, a
diffusion coefficient of 10$^{-3}$ cm$^2$/s at 400 K is given; for a
discussion, see Ref.\ \onlinecite{pablo}. The ``correct'' number, including
flux chopping, would be 1.0~cm$^2$/s {\em if monoclusters {\em only} were
assumed to be mobile}. However, as we have seen above, cluster dimers diffuse
at a rate which is quite comparable to that for monoclusters, suggesting that
larger clusters would diffuse as well. The Lennard-Jones simulations of
Deltour {\em et al.}\cite{del97} indicate that the rate of diffusion of {\it
compact} $N$-atom clusters scales roughly as the inverse of the contact area
between the cluster and the substrate: $D_{N} = D_1 N^{-2/3}$. (Compact
clusters are expected to form through aggregation and coalescence; see Ref.\
\onlinecite{pablo}).

Experimentally, however, it is almost impossible to determine whether or not
multiclusters do diffuse, and at which rate. In view of this, and the
expected importance of multicluster mobility on growth, we have carried out a
series of kinetic Monte Carlo (KMC) simulations in order to estimate the
largest island which must be allowed to diffuse in order to account for the
experimentally-observed gold island density on graphite at 400 K, viz.\
$4\times10^{8}$ islands/cm$^{2}$, or $1.1\times10^{-5}$ per
site.\cite{pablo,ss} To do so, we assume that the diffusion constant for
monoclusters found in the present simulations is correct, and that the rate
of diffusion of $N$-clusters scales according to the law given above. All
other parameters (incident cluster flux, temperature, chopping rate, etc.)
are fixed by experiment.

\begin{figure}
%\vspace*{1cm}
\vfill
\epsfxsize=8cm
\epsfbox{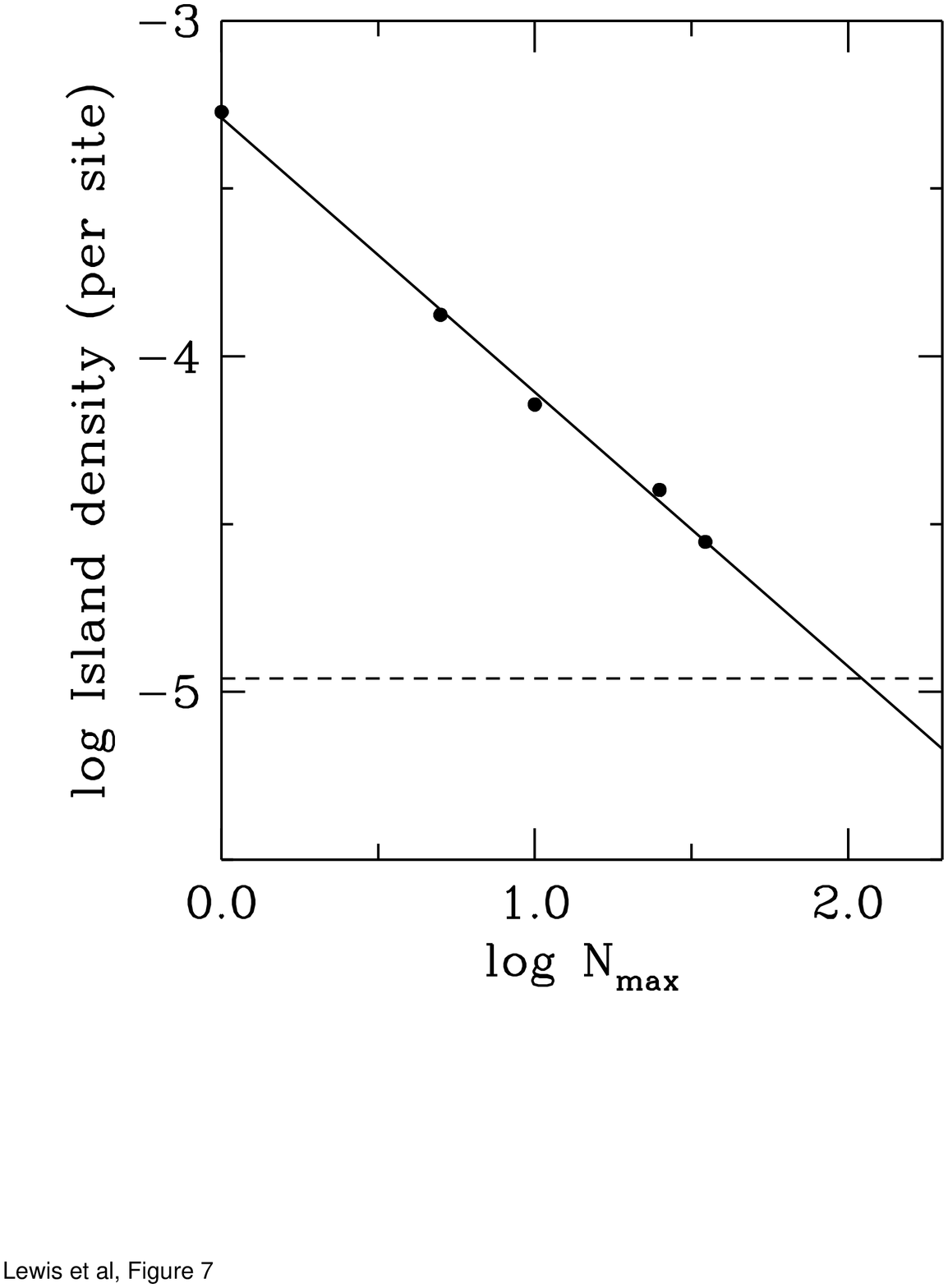}
\vspace*{0.5cm}
\caption{
Predicted island density as a function of the size $N_{\rm max}$ of the
largest multi-cluster island which is allowed to diffuse. The calculations
were carried out using kinetic Monte Carlo simulations as discussed in the
text. The full line is a linear fit to the data points and the dashed line
indicates the experimental density.
\label{fig_nmax}
}
\end{figure}

Figure \ref{fig_nmax} shows the results of the KMC simulations: we plot here
the island density that would be observed if the largest mobile island were
of size $N_{\rm max}$. The computational load increases very rapidly with
$N_{\rm max}$ and we therefore only considered islands of sizes less than or
equal to 35. The data points follow very closely a power-law relation and we
can thus extrapolate to larger values of $N_{\rm max}$, i.e., smaller island
densities. We find in this way that islands up to a maximum size of about 100
mono-clusters must be mobile in order to account for the observed island
density of $1.1\times10^{-5}$ per site. In what follows, we discuss in more
detail the connection of this observation with experiment.

We first note that, in the gold-on-gra\-phi\-te experiments,\cite{pablo}
large islands form which are ``partially ramified'', in the sense that the
branch width is much larger than the size of the deposited clusters, each
branch being formed by the coalescence of up to 200 mo\-no\-clus\-ters. In
contrast, for antimony cluster deposition on graphite at room
temperature,\cite{pablo,PJLB} the islands are fully ramified, i.e., have a
branch width identical to the diameter of the mo\-no\-clus\-ters; this
establishes unambiguously that cluster coalescence is not taking place in
this case. It has been shown, further, that the mobility of the islands is
negligible in antimony.\cite{ss} Our results suggest, therefore, when taken
together with the work of Deltour {\em et al.},\cite{del97} that {\em
compact} islands, which form through diffusion and coalescence, are mobile
according to a $N^{-2/3}$ law. In contrast, ramified islands, which form when
coalescence does not take place, have much reduced mobility -- certainly much
less than would be expected from a $N^{-2/3}$ law. $N_{\rm max}$, therefore,
signals the crossover point between the two mobility regimes or,
equivalently, the multicluster size at which the morphology of the islands
crosses over from compact to ramified (or vice-versa). The physical reasons
underlying the relation between mobility and morphology are not clear, but
there appears to be no other ways to interpret the experimental results. This
problem clearly deserves further studies.

To summarize this section, the mobility of large islands is evidently a {\em
necessary} ingredient to account for the experimentally observed island
density. Our simulations suggest that these islands can be as large as 100
monoclusters; while this is consistent with experiment, the exact value, as
well as the precise dependence of the diffusion rate on size, cannot at
present be estimated.

\section{Concluding remarks}
\label{concl}

Cluster-deposition techniques are of great potential interest for assembling
materials with specific, tailor-made applications. Yet, the fabrication
process depends critically on the possibility for the clusters to diffuse on
the surface in order to settle in appropriate positions, thus forming
self-organized structures, or to aggregate/coalesce with other clusters in
order to form larger-scale structures and eventually continuous layers. In
this article, we have demonstrated, using molecular-dynamics simulations with
realistic interatomic potentials, that the diffusion of large metallic
clusters on graphite can take place at a pace which is quite comparable to
that for single adatoms. We have also established that the rate of diffusion
of cluster dimers can be very sizeable, comparable in fact to that for
monoclusters. An extremely important consequence of this is that islands
formed by the aggregation of clusters are also expected to be mobile. Using
kinetic Monte Carlo simulations and assuming a proper scaling law for the
dependence on size of the diffusivity of large clusters, we estimate that
islands containing as much as 25~000 atoms (100 monoclusters) are expected to
undergo diffusion at a significant rate on graphite surfaces. These findings
have profound consequences for the morphology of cluster-assembled thin
films.

\acknowledgements

We are grateful to Laurent Bardotti, Art Voter and Tapio Ala-Nissila for
useful discussions. This work was supported by the Natural Sciences and
Engineering Research Council of Canada and the ``Fonds pour la formation de
chercheurs et l'aide \`a la recherche'' of the Province of Qu\'ebec. LJL is
grateful to the {\em D\'epartement de physique des mat\'eriaux de
l'Universit\'e Claude-Bernard-Lyon-I}, where part of this work was carried
out, for hospitality, support, and pleasant weather.

%\newpage

%\newpage

\vfill

\begin{figure}
\epsfxsize=8cm
\epsfbox{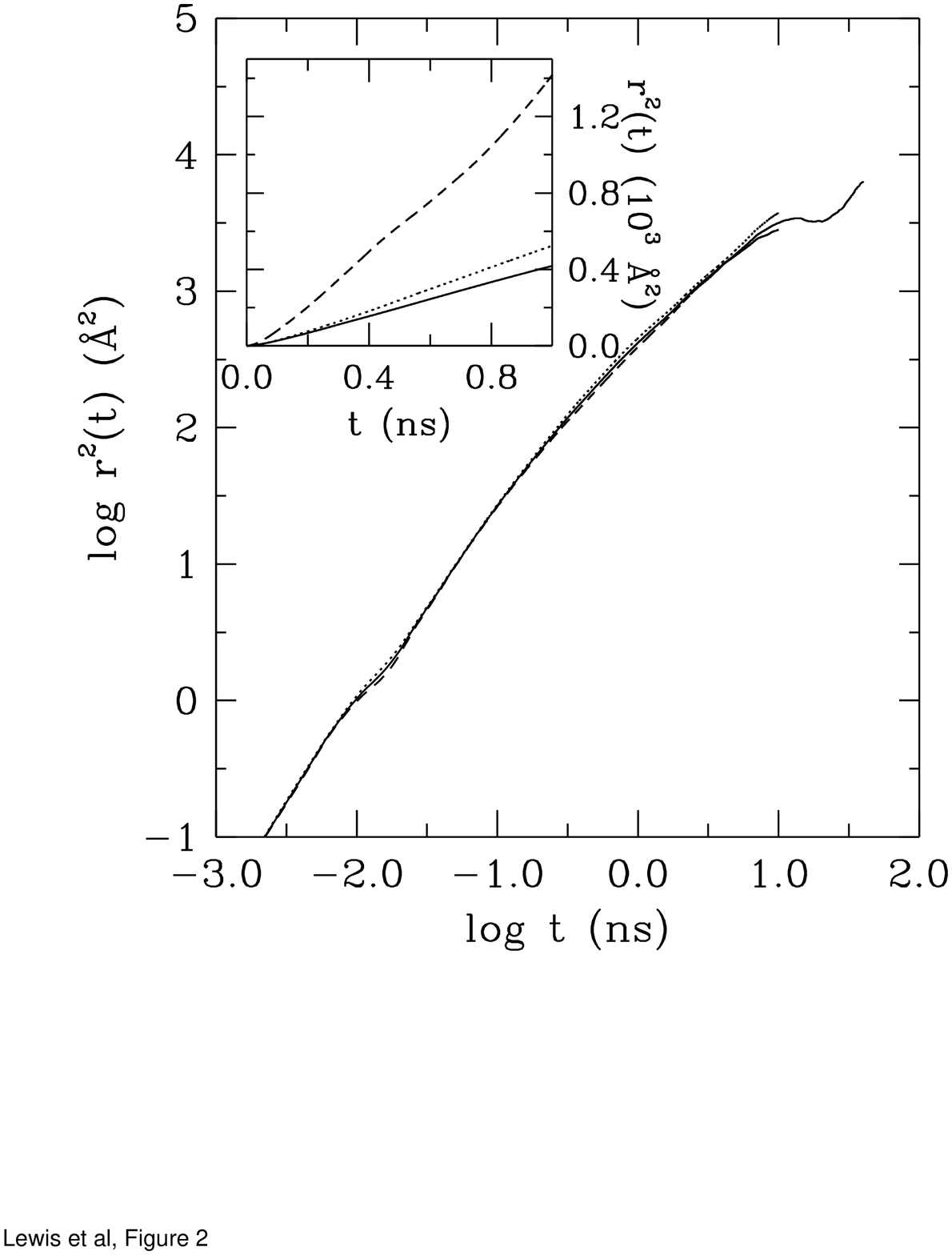}
%\vspace*{-5cm}
\caption{
Main figure: Log-log plot of the time-averaged mean-square displacements for
the cluster's center-of-mass on the static substrate at 500 K. The three
curves correspond to different estimates: using the full extent of the run
(full curve); only the first half (dashes); only the second half (dots). The
difference between these curves gives a measure of the error on the estimated
diffusion coefficient. Inset: Time-averaged mean-square displacements for the
monocluster on a static substrate (full line), the monocluster on a dynamic
substrate (dashes) and the dicluster on a static substrate (dots).
\label{fig_diff_stat}
}
\end{figure}

\begin{references}
\bibitem{note} Three different estimates of $r^2(t)$ are presented in Fig.\
\protect\ref{fig_diff_stat}: using the full extent of the run, only the first
half, and only the second half. They are almost undistinguishable on this
log-log plot.
\end{references}
\end{document}